\documentclass[%
 reprint,
superscriptaddress,
 amsmath,amssymb,
 aps,
 pra,
]{revtex4-2}

\usepackage{graphicx}%
\usepackage{dcolumn}%
\usepackage{bm}%
\usepackage{xcolor}
\usepackage[colorlinks=true,linkcolor=blue,urlcolor=blue,citecolor=blue]{hyperref}

\newcommand{\Eqref}[1]{\mbox{Eq.\hspace{0.25em}\eqref{#1}}}
\newcommand{\Eqsref}[1]{\mbox{Eqs.\hspace{0.25em}\eqref{#1}}}
\newcommand{\Figref}[1]{\mbox{Fig.\hspace{0.25em}\ref{#1}}}

\newcommand{\Nd}{N_\mathrm{D}}

\newcommand{\optionalsection}[1]{\section{#1}}
\newcommand{\optionalsubsection}[1]{\subsection{#1}}

\begin{document}

\title{Oscillating concentrations suppress condensate coarsening} %

\author{Mathias S. Heltberg}
\thanks{These authors contributed equally.}
\affiliation{Niels Bohr Institute, University of Copenhagen, Copenhagen, 2100, Denmark}

\author{Lukas H. Kristensen}%
\thanks{These authors contributed equally.}
\affiliation{Niels Bohr Institute, University of Copenhagen, Copenhagen, 2100, Denmark}

\author{Mogens H. Jensen}
\affiliation{Niels Bohr Institute, University of Copenhagen, Copenhagen, 2100, Denmark}

\author{David Zwicker}
\affiliation{Max Planck Institute for Dynamics and Self-Organization, Am Faßberg 17, 37077 Göttingen, Germany}%

\begin{abstract}
Living cells utilize condensates to spatially concentrate molecules in response to dynamic signals. For instance, nuclear condensates respond to oscillations in transcription factor levels in the nucleoplasm, including those involved in repairing multiple DNA breaks. To understand how oscillating signals affect condensates, we analyze a theoretical model using numerical simulations and analytical theory. While passive dynamics would drive all molecules into a single condensate, we find that sufficiently fast oscillations stabilize multiple droplets, allowing control of their sizes. We thus reveal a new behavior of chemically active droplets, which could be exploited in synthetic applications.

\end{abstract}

\maketitle

\optionalsection{\label{sec:level1}Introduction}

Living cells use biomolecular condensates to organize their interior in space and time, enabling functions such as gene regulation and DNA repair at specific intracellular locations~\cite{Banani2017,brangwynne2009germline,larson2017liquid,strom2017phase,kilic2019phase,Lasker2022,Zhang2021c,sabari2018coactivator,du2024direct,pessina2019functional,mine2021single,heltberg2022enhanced,chin2024parp1}. These membrane-less compartments typically form by phase separation driven by weak multivalent interactions~\cite{Hyman2014,zwicker2025physics}. However, phase separation generally implies surface tension and thus coarsening: droplets merge by coalescence and larger droplets grow at the expense of smaller ones by Ostwald ripening~\cite{lifshitz1961kinetics,wagner1961theorie,Voorhees1985}. Since many cellular processes require multiple condensates to coexist, cells must suppress coarsening. While several passive and active mechanisms for stabilizing droplets in static settings have been identified~\cite{webster2001osmotic,Fernandez-Rico2023,Qiang2023,vidal2021cavitation,Glotzer1994,Christensen1996,zwicker2015suppression,kirschbaum2021controlling,Luo2024a}, how condensates are controlled in the dynamically changing environment of living cells remains poorly understood.

A particularly intriguing aspect of cellular dynamics is that many regulatory proteins exhibit oscillatory concentrations on time scales of hours, including NF-$\kappa$B, Hes1, and p53~\cite{hoffmann2002ikappab,nelson2004oscillations,kobayashi2009cyclic,lahav2004dynamics}. Such oscillations are often assumed to play functional roles~\cite{heltberg2021tale}, but their physical consequences for condensates remain largely unexplored. In this work, we show that oscillations in the concentration of droplet material can suppress Ostwald ripening and stabilize multiple droplets. To this end, we consider nonequilibrium addition and degradation of droplet material, which prevent global equilibration and thereby arrest coarsening. Our results yield testable predictions for how oscillation amplitude and frequency affect droplet stability and suggest a possible physical function of protein oscillations in controlling intracellular condensates.

\optionalsection{Results}
\optionalsubsection{Oscillations suppress Ostwald ripening}
To study droplets in an oscillating environment, we consider a system of fixed volume~$V_\mathrm{sys}$ comprising a binary mixture of droplet material representing a segregating protein, such as p53, and a second species, which represents the many different molecules in the nucleoplasm.
The system's state is then characterized by the concentration~$c(\vec x, t)$ of droplet material at position $\vec x$ and time~$t$.
For simplicity, we assume equal molecular volumes~$\nu$ for all species, so that the remaining nucleoplasm has concentration $c_{\mathrm{nuc}}(\vec{x},t) = \nu^{-1}-c(\vec{x}, t)$. 
We describe phase separation by a Ginzburg--Landau free energy density
\begin{equation}
    h(c) = -\frac{b}{2}\!\left(c - \frac{1}{2\nu}\right)^2 
    + \frac{a}{4}\!\left(c - \frac{1}{2\nu}\right)^4
    \label{eqn:free_energy_density}
\end{equation}
with positive coefficients $a$ and $b$, which set the coexisting compositions $c_\mathrm{in} = (2\nu)^{-1}+\sqrt{b/a}$ and $c_\mathrm{out} = (2\nu)^{-1}-\sqrt{b/a}$ of thermodynamically large phases.
To describe finite-sized droplets, we account for spatial variations by introducing the free energy functional
\begin{equation}
    H[c] = \int_{V_\mathrm{sys}} 
    \left[ h(c) + \frac{\kappa}{2}\, |\nabla c|^2 \right] \, \mathrm{d}V,
    \;,
\label{eq:free_energy}
\end{equation}
where the gradient penalty~$\kappa$ controls interfacial width $\eta = \sqrt{2\kappa/b}$ and surface tension $\gamma=2\sqrt{2\kappa b^3}/3a$ \citep{weber2019physics,zwicker2025physics}.
Passive systems then evolve by minimizing the free energy $H$, which leads to droplet coarsening to minimize the overall interfacial area~\cite{Voorhees1985}.

\begin{figure}
\centering
	\includegraphics[width=\linewidth]{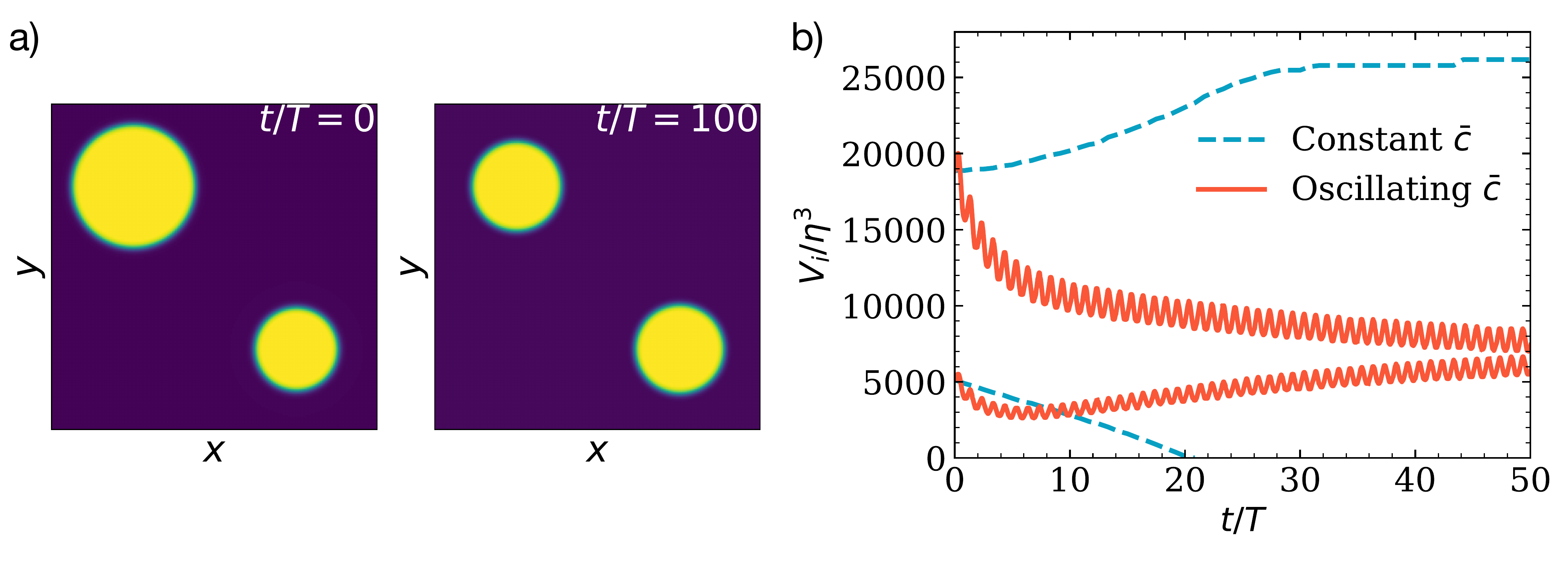}
	\caption[]{
        (a) Maximum projection of concentration field $c(\vec x)$ from 3D simulations of \Eqref{eq:cahn_hilliard} for two indicated times $t$ and $c_0\approx 0.18 \frac1\nu$, $c_\mathrm{in}-c_\mathrm{out}\approx0.71 \frac1\nu$,
        $T \approx 84 \eta^2/\Lambda a$, $A \approx 0.018 (c_\mathrm{in} - c_\mathrm{out})$,
         and  $V_\mathrm{sys}\approx8.5 \cdot 10^5 \eta^3$. %
        (b)~Droplet volumes $V_i$ of two droplets as a function of $t$ for oscillating $\bar c$ (red solid lines, corresponding to panel a) and constant $\bar c$ (blue dashed lines).
    }
    \label{Fig_phase_field}
\end{figure}

We next introduce oscillations, which we conceptualize as production and degradation of droplet material.
We describe this situation by the dynamical equation
\begin{align}
    \frac{\partial c(\vec{x}, t)}{\partial t} = \Lambda \nabla^2 \frac{\delta H}{\delta c} + s(c, t)
    \;, 
\label{eq:cahn_hilliard}
\end{align}
where the first term on the right hand side describes passive, diffusive fluxes proportional to gradients in chemical potential $\delta H/\delta c$ and the constant mobility~$\Lambda$.
In contrast, the reaction flux~$s$ accounts for production and degradation leading to an oscillating volume-averaged concentration $\bar c = V_\mathrm{sys}^{-1}\int c \, \mathrm{d} V$ of droplet material.
For simplicity, we model these oscillations by a sinusoidal wave form with angular frequency $\omega$ and amplitude $A$ around a time averaged value~$c_0$,
\begin{align}
    \bar{c}(t) = c_0 + A\sin(\omega t)
    \;.
    \label{eq:concentration_oscillations}
\end{align}
To achieve this average behavior, we prescribe a simple protocol for production and degradation, where the reaction flux~$s$ depends linearly on $c$, implying $\frac{\mathrm{d}\bar{c}}{\mathrm{d}t} = s(\bar c, t)$.
In particular, we assume that droplet material is converted to nucleoplasm ($\frac{\mathrm{d}\bar{c}}{\mathrm{d}t} < 0$) proportional to the concentration of droplet material, e.g., because all molecules are degraded at the same rate, regardless of phase.
Equivalently, we assume that droplet material is created from nucleoplasm at an equal rate for $\frac{\mathrm{d}\bar{c}}{\mathrm{d}t} > 0$. %
Hence,
\begin{equation}
    s(c, t) = 
    \dfrac{\mathrm{d}\bar{c}}{\mathrm{d}t} \cdot \begin{cases}
    \dfrac{\nu^{-1}-c(\vec{x}, t)}{ \nu^{-1} - \bar{c}(t)} & \text{if } \quad  \dfrac{\mathrm{d}\bar{c}}{\mathrm{d}t} > 0\; \\[12pt]
     \dfrac{c(\vec{x}, t)}{\bar{c}(t)}  & \text{if } \quad  \dfrac{\mathrm{d}\bar{c}}{\mathrm{d}t} < 0 \;.
    \end{cases}
    \label{eq:SReaction}
\end{equation}
Fundamentally, this implies that conversion of nucleoplasm to droplet material happens at a higher rate in the dilute phase, whereas the converse conversion is dominant in the dense droplet phase.
Numerical simulations of the model given by \Eqsref{eqn:free_energy_density}--\eqref{eq:SReaction} reveal that two prepared droplets evolve toward the same size (\Figref{Fig_phase_field}a), contrary to the expected Ostwald ripening in passive systems.
The individual droplet volumes oscillate with the imposed frequency~$\omega$ and converge toward the same value on much longer time scales (\Figref{Fig_phase_field}b).

\optionalsubsection{Suppression of ripening requires minimal amplitude and frequency}

To understand the observed suppression of Ostwald ripening, we next quantify how the amplitude $A$ and frequency $\omega$ affect the dynamics of droplet sizes.
To simplify the equations, we assume that the $\Nd$ droplets in the system maintain a spherical shape described by radii $R_i$ and exhibit a thin interface ($\eta\ll R_i$) for $i=1,\ldots, \Nd$. %
Assuming droplets evolve quasistatically and concentration fields in their vicinity are spherically symmetric, the droplet radii~$R_i$ are the only dynamical degrees of freedom~\cite{zwicker2025physics}, akin to Lifshitz-Slyozov-Wagner theory~\cite{lifshitz1961kinetics,wagner1961theorie}.
The volumes $V_i=\frac{4}{3}\pi R_i^3$ then evolve according to~\cite{weber2019physics}
\begin{align}
    \frac{\mathrm{d}V_i}{\mathrm{d}t} = \frac{4\pi \hat{D}}{c_\mathrm{in}}  \bigg[ V_i^{\frac{1}{3}} \bigl(c_\infty(t)-c_\mathrm{out}\bigr)- \hat{\ell}_\gamma c_\mathrm{out} \bigg] + S^\mathrm{in}_i(t)
    \;,
    \label{eq:DropletGrowth}
\end{align}
where $S^\mathrm{in}_i(t) = c_\mathrm{in}^{-1}V_i(t) \, s(c_\mathrm{in}, t)$ describes the direct effect of the oscillations, which follow from \Eqref{eq:SReaction} assuming a homogeneous profile $c(\vec{x}, t) = c_\mathrm{in}$ in the limit of large reaction-diffusion length scales ($\sqrt{D/\omega} \gg R_i$).
In contrast, the first term in \Eqref{eq:DropletGrowth} accounts for growth due to diffusive fluxes with effective diffusivity $\hat{D} = D [ \frac{3}{4\pi} ]^{1/3}$ using the diffusion constant $D = 2b \Lambda$, and a rescaled capillary length $\hat{\ell}_\gamma = \ell_{\gamma} [ \frac{4 \pi}{3} ]^{1/3}$, where $\ell_\gamma\approx 2\gamma/(c_\mathrm{in}k_BT)$.
Apart from surface tension effects described by $\ell_\gamma$, the droplet grows if the equilibrium concentration $c_\mathrm{in}$ is smaller than the concentration~$c_\infty(t)$ far away, which follows from material conservation,
\begin{align}
    c_\infty(t) = \frac{\bar{c}(t) V_\mathrm{sys} - c_\mathrm{in}\sum_{i=1}^{\Nd} V_{i}}{V_\mathrm{sys}-\sum_{i=1}^{\Nd} V_{i}}
    \;.
    \label{eq:cInf}
\end{align}
Consequently, oscillations of the average concentration $\bar c$ imply oscillating $c_\infty$, which affects droplet growth.

\begin{figure}
\centering
	\includegraphics[width=\linewidth]{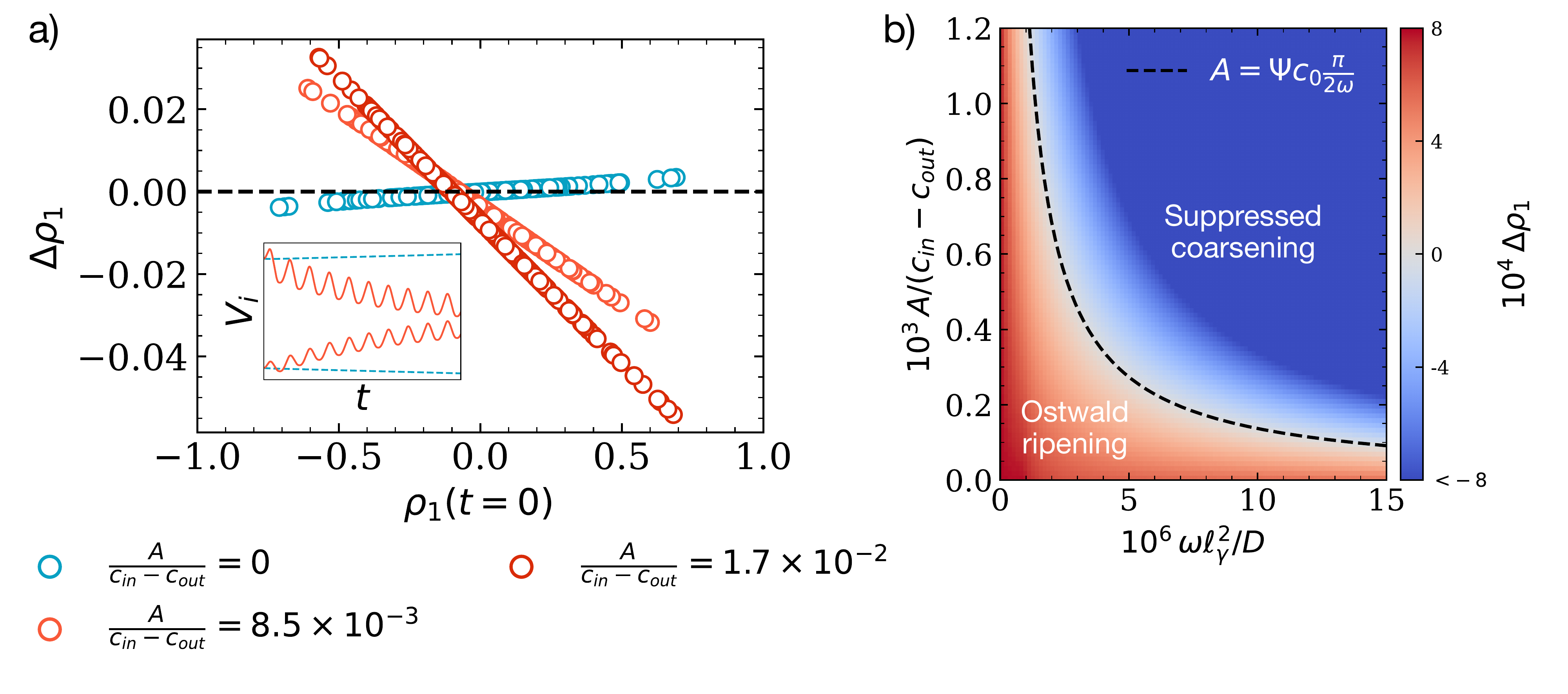}
	\caption[]{
        (a) Variation $\Delta \rho_1 = \rho_1(T)-\rho_1(0)$ of relative droplet volumes $\rho_1 = (V_1 - V_2)/(V_1 + V_2)$ over one period~$T$ as a function of $\rho_1$ for various amplitudes~$A$ %
        averaged over 200 simulations of \Eqref{eq:DropletGrowth} with random initial conditions. %
        Inset: examples of volumes $V_i$ as a function of time $t$. %
        Parameters are $V_\mathrm{sys} \approx 3.35 \cdot 10^7 \ell_{\gamma}^3$, $c_0 \approx 0.085 (c_\mathrm{in}-c_\mathrm{out})$, and $T \approx 1.3 \cdot 10^4 \ell_{\gamma}^2/D$.
        (b) $\Delta \rho_1$ as a function of oscillation frequency $\omega$ and amplitude $A$.
        Coarsening is suppressed if $\Delta\rho_1<0$ (blue region), consistent with \Eqref{eq:omegaArelation} (black dashed line).
        Model parameters are $V_\mathrm{sys} = 1.25 \cdot 10^7 \ell_{\gamma}^3$ and $c_0 \approx 0.21 (c_\mathrm{in} - c_\mathrm{out})$.}
    \label{fig:ThinInterfaceApp}
\end{figure}

To see whether the approximations introduced above are reasonable, we solve \Eqsref{eq:DropletGrowth}--\eqref{eq:cInf} numerical.
The effective model reproduces our initial observations: Oscillations enable two droplets to converge to a common volume, whereas the smaller droplet dissolves in a passive system.
We quantify these dynamics by introducing the volume deviation
$\delta V_{i}(t) = V_{i}(t) - \bar{V}(t)$,
where $\bar V(t)=\Nd^{-1}\sum_{i=1}^{\Nd} V_{i}(t)$ is the time-dependent mean volume, implying $\sum_{i=1}^{\Nd} \delta V_{i}(t) = 0$.
We then measure relative deviations $\rho_i(t) = \delta V_i(t)/\bar{V}(t)$, which are positive for droplets larger than $\bar V$.
Quantifying the relative growth $\Delta \rho_i = \rho_i(T)-\rho_i(0)$ over one period $T=2\pi/\omega$, we observe that large droplets ($\rho_i>0$) grow ($\Delta\rho_i>0$) in passive systems (blue), whereas they shrink ($\Delta\rho_i<0$) in sufficiently strongly driven systems  (\Figref{fig:ThinInterfaceApp}a).
Interestingly, $\Delta\rho_i$ depends linearly on the initial size $\rho_i$ in all cases, suggesting exponential dynamics.

To quantify the influence of the amplitude~$A$ and frequency~$\omega$ of the oscillation onto the droplet dynamics, we next analyze a droplet pair at fixed $\rho_i(0)$ and measure the corresponding change in sizes.
\Figref{fig:ThinInterfaceApp}b shows that both amplitude and frequency must exceed critical values to suppress Ostwald ripening.
The line of stability is approximated by an inverse proportional relationship, suggesting that the product $\omega A$ is a fundamental control parameter.
Since $\omega A$ governs the production and degradation rate, $\frac{\mathrm{d}\bar{c}}{\mathrm{d}t} =\omega A \sin(\omega t)$, we hypothesize reactions need to be fast enough to suppress Ostwald ripening.

\optionalsubsection{Asymmetric growth and shrinkage stabilize droplet sizes}
\label{section:analytical}

\begin{figure}
    \centering
	\includegraphics[width=\linewidth]{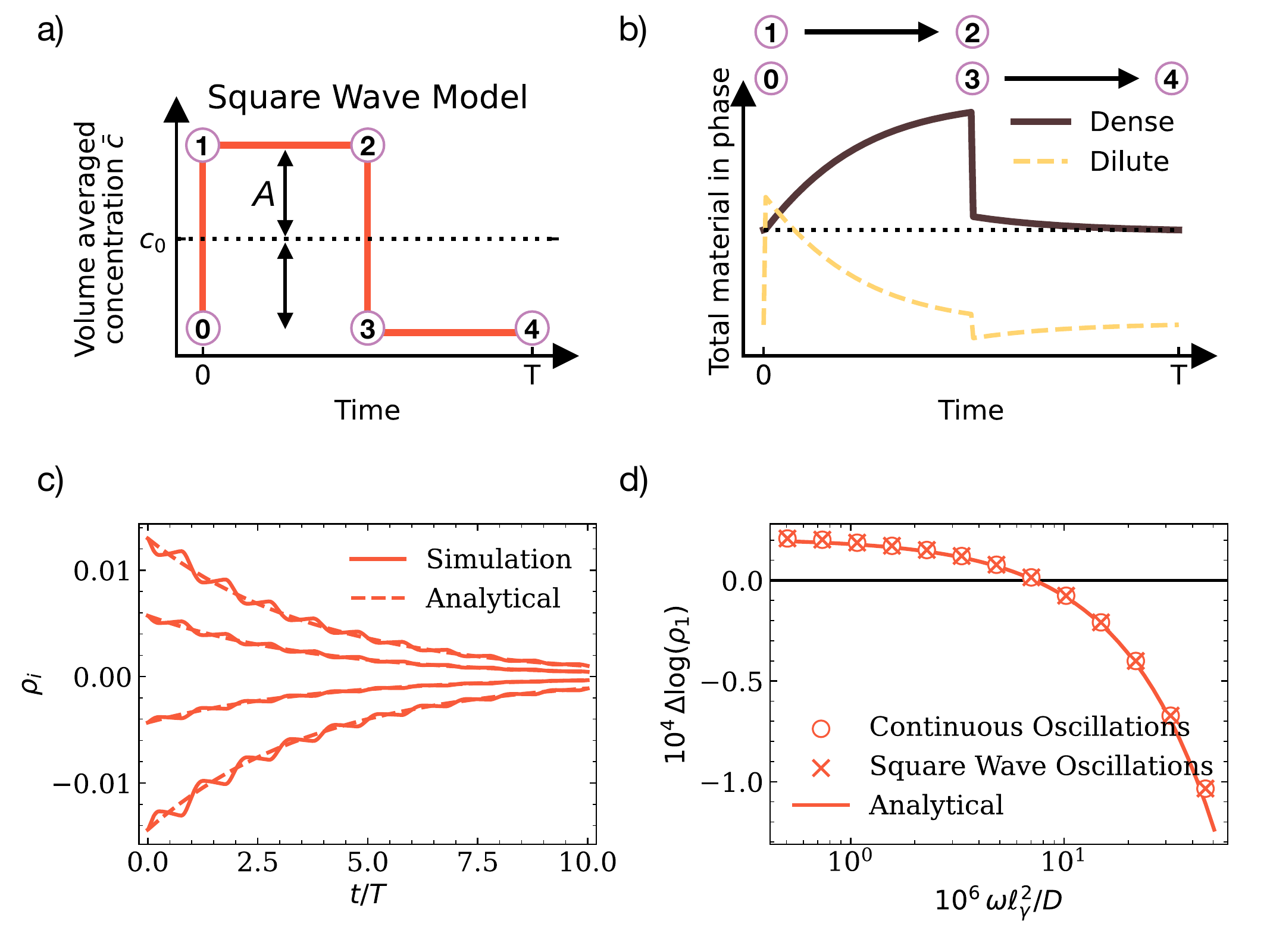}
	\caption[]{
        (a) Schematic representation of square wave model. 
        (b) Total amount of droplet material in dense and dilute phases as a function of time~$t$ over one period~$T$.
        (c) Relative volumes $\rho_i = V_i/\bar V - 1$ for four droplets as a function of $t$ comparing simulations (solid lines) to analytical result of \Eqref{eq:RhoCycle_SS} (dashed lines).
        Parameters are $A \approx 0.042 (c_\mathrm{in} - c_\mathrm{out})$, $T = 9.5 \cdot 10^5 \ell_{\gamma}^2/D$, $V_\mathrm{sys} = 1.0\cdot 10^8 \ell_{\gamma}^3$, and $c_0 \approx 0.085 (c_\mathrm{in}-c_\mathrm{out})$.
        (d) Variation of $\rho_1$ over one period, $\Delta \log(\rho_1) = \log[\rho_1(T)] - \log[\rho_1(0)]$, as a function of oscillation frequency~$\omega$ for harmonic oscillations (circles) and square wave oscillations (crosses), compared to the analytical results (\Eqref{eq:RhoCycle_SS}, line).
        Parameters are $V_\mathrm{sys} = 1.6 \cdot 10^{9} \ell_\gamma^3$, $c_0 \approx 0.26 (c_\mathrm{in} - c_\mathrm{out})$, and $A = 0.018 (c_\mathrm{in} - c_\mathrm{out})$.
        }
    \label{fig:analytical}
\end{figure}

To understand the mechanism of the suppression of Ostwald ripening by oscillatory reactions, we next analyze in detail how material is shared between several droplets.
To do this, we approximate the sinusoidal oscillations described by \Eqref{eq:concentration_oscillations} by a square wave that oscillates the average composition $\bar c$ between $c_0-A$ and $c_0+A$.
This simplification reduces a single cycle to a process involving four stages, separating reactions from diffusive transport (\Figref{fig:analytical}a).
In particular, all reactions are instantaneous, with nucleoplasm being converted to droplet material between points 0--1, and droplet material being converted to nucleoplasm between points 2--3, implying that the spatially averaged concentration $\bar c^{(n)}$ at point~$n$ is given by $\bar c^{(0)}=\bar c^{(3)}=\bar c^{(4)}=c_0 - A$ and $\bar c^{(1)}=\bar c^{(2)}=c_0 + A$.
Between these points, i.e., during stages 1--2 and 3--4, reactions are absent and material is merely transported between phases (\Figref{fig:analytical}b).

To analyze the dynamics of the square wave model, we determine the droplet volumes $V_i^{(n)}$ at the various points $n=0,\ldots, 4$.
Starting with an initial volume $V_i^{(0)}$, the production of droplet material will lead to an increase in stage 0--1, $V_i^{(1)} = \Gamma_i^+ V_i^{(0)}$ with $\Gamma_i^+\ge 1$.
The limit $\Gamma_i^+=1$ is reached in the simple case of strong phase separation where the droplet hardly contains any nucleoplasm and production inside the droplet is thus negligible.
Conversely, the degradation of droplet material in stage 2--3 necessarily affects the droplet volume, $V_i^{(3)}=\Gamma_i^- V_i^{(2)}$ with $\Gamma_i^-<1$.
Applying the logic of \Eqref{eq:SReaction} to the square wave model, we have
$V_i^{(3)}-V_i^{(2)} 
= V_i^{(2)}\frac{-2A}{c_0 + A}$, implying 
$\Gamma_i^- =(c_0-A)/(c_0+A)$.
Note that both reactions change all droplet volumes by the same factors $\Gamma_i^\pm$, independent of the droplet identity $i$.
Consequently, the relative droplet volumes $\rho_i = V_i \bar V^{-1} - 1$ do not change during stages 0--1 and 2--3, so ripening must be suppressed by diffusive exchange during stages 1--2 and 3--4.

To understand how diffusive transport between droplets suppresses Ostwald ripening, we next analyze stages 1--2 and 3--4, where $S^\mathrm{in}_i(t) = 0$.
Each stage comprises two intertwined dynamics:
First, the average volume~$\bar V$ exhibits a near-exponential relaxation to the new target volume set by the background concentration in the stage (\Figref{fig:analytical}b).
Second, each relative droplet volume~$\rho_i$ reacts to the evolving value of $\bar V$.
Assuming small deviations and surface tension effects, the dynamics during stage 1--2 imply (Appendix~\ref{sec:appendix_square}) 
\begin{align}
    \rho_i^{(2)} 
    &\approx 
    \rho_i^{(1)} 
    \exp\!\left(\frac{4\pi^2 D l_\gamma c_\mathrm{out}}{3\omega c_\mathrm{in}}\, 
    \frac{1}{\bar{V}^{+}_\mathrm{SS}} \right)
    \left( \frac{\bar{V}^{(2)}}{\bar{V}^{(1)}} \right)^{-\frac{2}{3}}
    \;,
    \label{eq:rho2}
\end{align}
where $\bar{V}^\pm_\mathrm{SS} \approx V_\mathrm{sys}\Nd^{-1}  (c_0 \pm A -c_\mathrm{out})/(c_\mathrm{in}-c_\mathrm{out})$.
Here, the argument of the exponential is positive, so that the corresponding factor increases the magnitudes of the relative volumes $\rho_i$, which implies coarsening.
In contrast, the last factor is smaller than one and thus related to the suppression of coarsening during stage 1--2.
We obtain a similar expression for stage 3--4. %

Finally, we combine our results to express the change of the relative droplet volumes~$\rho_i$ over one cycle.
Since $\rho_i$ does not change during the reaction stages, we obtain the final value $\rho_i^{(4)}$ from the initial value $\rho_i^{(0)}$ by first using \Eqref{eq:rho2} and then the equivalent expression for stage \mbox{3--4}.
To simplify the result, we further assume $A \ll c_0$ to approximate $(\bar V_\mathrm{SS}^+)^{-1} + (\bar V_\mathrm{SS}^-)^{-1} \approx 2 (V_\mathrm{SS}^0)^{-1}$, where $\bar{V}^0_\mathrm{SS} = V_\mathrm{sys}\Nd^{-1}  (c_0 -c_\mathrm{out})/(c_\mathrm{in}-c_\mathrm{out})$ is the stationary mean droplet size without oscillations.
Taken together,
\begin{multline}
    \frac{\rho_i^{(4)}}{\rho_{i}^{(0)}} 
    \approx 
       e^{\frac{2\pi}{3\omega}\Psi}
        \left(\frac{\bar{V}^{(2)}}{\bar{V}^{(0)}\Gamma^+}\frac{\bar{V}^{(4)}}{\bar{V}^{(2)}\Gamma^-} \right)^{-\frac{2}{3}}
    = 
    e^{\frac{2\pi}{3\omega}\Psi} \bigl(\Gamma^+\Gamma^- \bigr)^{\frac{2}{3}}
    \\[4pt]
    \text{with } \Psi = \frac{4\pi D \ell_\gamma \Nd}{V_\mathrm{sys}}\frac{c_\mathrm{out}}{c_\mathrm{in}} \frac{c_\mathrm{in}-c_\mathrm{out}}{c_0 - c_\mathrm{out}}
    \;,
    \label{eq:Rhostabilization}
\end{multline}
where the final equality on the first line assumes mean droplet volumes to return to steady state ($\bar{V}^{(4)} = \bar{V}^{(0)}$).
\Eqref{eq:Rhostabilization} again separates dynamics due to coarsening (captured by the coarsening rate $\Psi$) from its suppression (captured by the product $\Gamma^+\Gamma^-$).
In particular, oscillations opposed coarsening if $\Gamma^+\Gamma^- < 1$, which is typically the case for square waves with strong phase separation ($\Gamma^+=1$).
In any case, this results shows that coarsening can only be suppressed if the degradation reaction affects droplets directly ($\Gamma^- < 1$); Merely affecting the surroundings and the associated diffusive exchange is insufficient.

To extend our results to sinusoidal oscillations, we apply the ideas presented in the previous paragraphs in infinitesimal steps.
In particular, we decompose the oscillations described by \Eqref{eq:concentration_oscillations} into separate reaction and diffusion steps. %
The change of the relative volume $\rho_i$ is then described by an equation akin to \Eqref{eq:rho2}, and we can determine the total change over one period~$T$ using Volterra's product integral (Appendix~\ref{sec:appendix_sine}),
\begin{align}
    \frac{\rho_i(T)}{\rho_{i}(0)} \approx e^{\frac{2\pi}{3\omega}\Psi} \left(e^{\Gamma^+-1} \frac{c_0-A}{c_0+A}\right)^{\frac{2}{3}}
    \;.
    \label{eq:RhoCycle_SS}
\end{align}
This relation holds for all droplets $i$, revealing that the relative change in $\rho_i$ for each droplet is constant and given by the parameters of the system.
This qualitative behavior is consistent with the relation $\Delta \rho_i \propto \rho_i$ that we obtained numerically (\Figref{fig:ThinInterfaceApp}a).
To test \Eqref{eq:RhoCycle_SS} quantitatively, we simulate a system of $\Nd=4$ droplets, and find that the stabilization rates are consistent with the prediction (Fig.~\ref{fig:analytical}c).
Moreover, we have $e^{\Gamma^+-1} \frac{c_0-A}{c_0+A} \approx \Gamma^+\Gamma^-$ for small amplitudes $A$, implying that \Eqref{eq:RhoCycle_SS} for harmonic oscillations and \Eqref{eq:Rhostabilization} for square oscillations are identical for small $A$.
Indeed, stabilization rates agree between the analytical result and simulations for both oscillation models  for a large range of frequencies $\omega$ (Fig.~\ref{fig:analytical}d).

The numerical simulations shown in \Figref{fig:ThinInterfaceApp}c suggest that coarsening is suppressed for sufficiently large reaction rates $\omega A$.
To understand this, we next ask when \Eqref{eq:RhoCycle_SS} predicts that $\Nd$ droplets are stable.
Using $\Gamma^+=1$ and assuming small $A$, so $(c_0-A)/(c_0+A) \approx e^{-\frac{2A}{c_0}}$, the right hand side of \Eqref{eq:RhoCycle_SS} is negative (so droplets are stable) when
\begin{align}
 \omega A > \frac{\pi}{2}\Psi c_0
 \;,
 \label{eq:omegaArelation}
\end{align}
with $\Psi$ defined in \Eqref{eq:Rhostabilization}.
This prediction quantitatively matches the numerically observed hyperbolic relationship between amplitude~$A$ and frequency~$\omega$ (\Figref{fig:ThinInterfaceApp}b).
In essence, the left hand side of \Eqref{eq:omegaArelation} summarizes the stabilizing effects of the oscillating drive, whereas the right hand side describes the coarsening process of Ostwald ripening driven by surface tension~$\gamma$ since $\Psi\propto \gamma$.

Our analysis of the square wave model reveals that oscillations stabilize multiple droplets by an asymmetry between growth and shrinkage:
When droplet material is added, all droplets grow by taking up material from the dilute phase, but the relative growth is larger in smaller droplets.
This is quantified by \Eqref{eq:rho2}, which implies $|\rho_i^{(2)}|<|\rho_i^{(1)}|$ for $\bar{V}^{(2)}>\bar{V}^{(1)}$.
These dynamics reduce the differences in relative sizes of droplets, thus stabilizing the system.
In contrast, the removal of material shrinks all droplets, and smaller droplets shrink faster than larger ones.
Since these dynamics mirror the growth phase, the stabilizing effect of growth is somewhat compensated.
However, material is also removed from droplets themselves (without changing $\rho_i$), so that over one cycle more material is added to the dilute phase than is removed from it.
This asymmetry in growth and shrinkage essentially suppresses coarsening.

\optionalsubsection{Faster oscillations imply smaller droplets}
\label{section:SingleDroplet}
Our results suggest that fast oscillations (larger $\omega$) stabilize droplets.
However, faster oscillations might also affect the average droplet size and droplet count $\Nd$, which influences $\Psi$.
In fact, the oscillatory drive can suppress droplet growth, even for individual droplets in the absence of coarsening.
To see this, consider the limit of an infinitely fast oscillation, which does not allow any diffusive exchange between phases.
Instead, droplets grow by $\Gamma^+$ when material is added and then shrink by $\Gamma^-$ when material is removed.
In the typical case where $\Gamma^+\Gamma^-<1$, droplets thus shrink continuously, implying that oscillations can suppress growth.

\begin{figure}
    \centering
	\includegraphics[width=\linewidth]{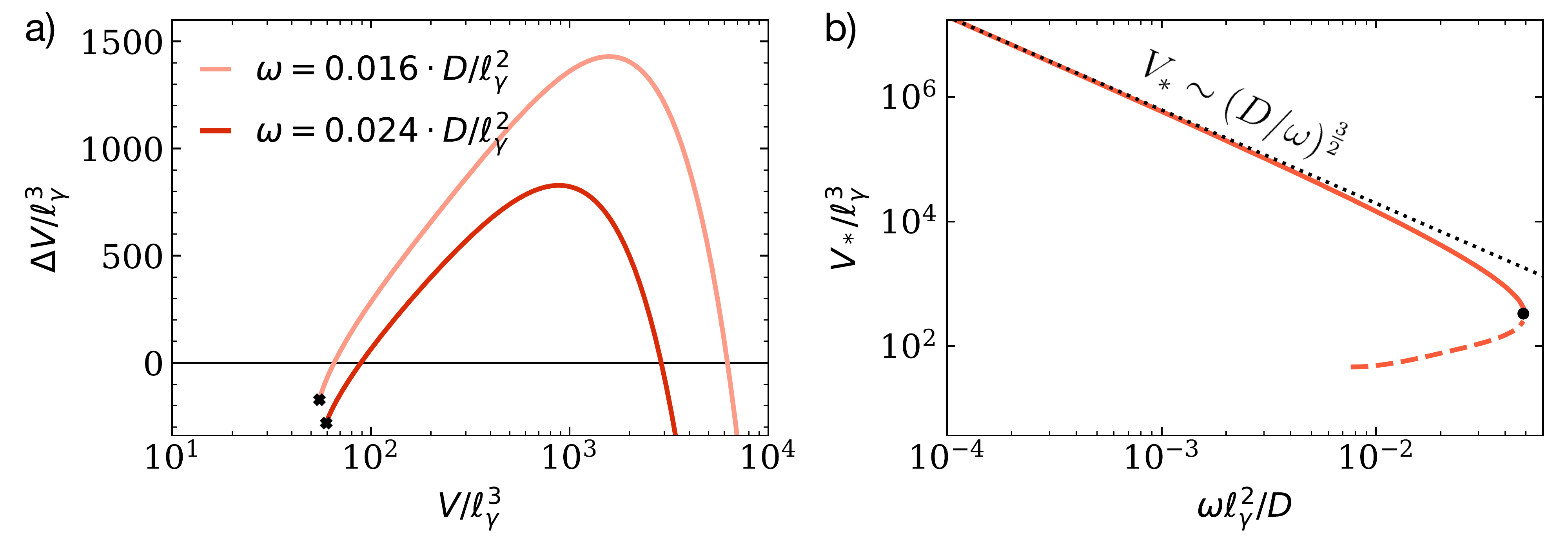}
	\caption{
    (a) Change $\Delta V = V(T)-V(0)$ of droplet volume~$V$ over one period as a function of $V$ for two oscillation frequencies~$\omega$ determined from solving \Eqref{eq:DropletGrowth}.
    Droplets dissolve fully within one period for $\omega$ smaller than a threshold (black crosses).
    (b) Stationary droplet volumes $V_*$ as a function of $\omega$, which are either stable (solid line) or unstable (dashed line).
    The dotted, black line indicates the theoretical scaling $V_* \propto (D/\omega)^{3/2}$. 
    (a--b) Parameters are $A \approx 0.018 (c_\mathrm{in}-c_\mathrm{out})$ and $c_0 \approx 0.25 (c_\mathrm{in}-c_\mathrm{out})$}
    \label{fig:single_droplet}
\end{figure}

Oscillations generally lead to a negative growth rate for sufficiently large droplets (\Figref{fig:single_droplet}a). 
Conversely, very small droplets also shrink due to surface tension effects.
However, if the frequency~$\omega$ is not too large, droplets of a finite range of volumes can grow.
This behavior implies two stationary states, of which the smaller one corresponds to the typical critical size also present in passive systems.
In contrast, the larger, stable stationary volume~$V^*$ corresponds to a droplet that does not change its size after one period.
Numerical simulations indicate that both stationary states depend on the frequency~$\omega$ and eventually meet in a saddle node bifurcation (\Figref{fig:single_droplet}b).
This implies that large frequencies cannot sustain any droplets, consistent with our simple argument in the previous paragraph.

In large system, where mass balance does not constrain droplet sizes, the steady droplet radius~$R_*$ must be controlled by the diffusive timescale in the droplet, $R_*^2/D$, and the relevant reactive time scale $\omega^{-1}$.
Balancing the two time scale predicts $R_* \sim (D/\omega)^{\frac{1}{2}}$, which is consistent with the observed scaling $V_* \sim (D/\omega)^\frac{3}{2}$ (\Figref{fig:single_droplet}b).
However, this simple argument breaks down for small droplets, where the interfacial width~$\eta$ becomes relevant, akin to similar behavior in chemically active droplets~\cite{zwicker2015suppression}.

\optionalsection{Discussion}

In summary, we have shown that oscillations in the amount of droplet material can suppress coarsening, thereby stabilizing multiple droplets and controlling their size. Since concentration oscillations are widespread in living systems, this mechanism may provide a generic route by which cells regulate condensate organization, for instance to parallelize biochemical processes such as DNA repair~\cite{heltberg2022enhanced} or to select a finite subset of droplets during developmental transitions~\cite{Cavka2025}. Mechanistically, suppression arises because cyclic driving removes material from droplets during the shrinkage phase and replenishes the dilute phase during growth, effectively redistributing material from larger to smaller droplets, akin to reaction-driven stabilization in chemically maintained emulsions~\cite{zwicker2015suppression,weber2019physics,zwicker2025physics}. Since this mechanism is robust, it should extend beyond the minimal model studied here to more complex multicomponent and dense-droplet systems, and it suggests new opportunities to probe condensate regulation \textit{in vivo}, \textit{in vitro}, and in synthetic soft-matter systems~\cite{Fritsch2021,Slootbeek2022,Peng2015}. More broadly, our results highlight that understanding condensates in dynamic cellular environments remains an open frontier, particularly when oscillations and phase separation become coupled through underlying reaction networks~\cite{haugerud2025excitability,sastre2025size,Bauermann2025a,Aierken2026}.

\textbf{Acknowledgements:}
We thank Alessandra Luchetti for discussions and critical reading of the manuscript.
MHJ acknowledges Novo-Nordisk Foundation grants no. NNF20OC0064978 and NNF24OC0089788.
MSH acknowledges Novo-Nordisk Foundation grant no. NNF23OC0085907.
MSH acknowledge the ERC Horizon Grant number 101221324 (PHOSCIL).
MHJ and LHK acknowledges the Carlsberg Foundation grant no. CF22-0494.
DZ acknowledges funding from the Max Planck Society and the European Research Council (ERC, EmulSim, 101044662).

\bibliography{bibliography}%

\clearpage
\appendix

\onecolumngrid

\section{Stability criterion for square wave forms}
\label{sec:appendix_square}
In this section, we derive the stability criterion $\omega A > \frac{\pi}{2}\Psi c_0$, given by \Eqref{eq:omegaArelation} in the main text by considering oscillations with square-wave forcing.
Perturbations in the fraction of droplet material are treated via a Taylor expansion around the resulting quasi-steady states.
Assuming small amplitudes, we linearize the dynamical equations and first solve for the mean droplet volume, $\bar{V}(t)$.
This solution is then used to compute the evolution of the relative deviation, $\rho_i$, over a full cycle.
Finally, we use these dynamics to derive the stability condition in the long-term dynamics, where the mean droplet volumes do not change over one cycle.

We describe the volume~$V_i$ of each of the $\Nd$ droplets based on their difference to the mean $\bar V$,
\begin{align}
	V_i = \bar{V}+ \delta V_i \quad \text{with} \quad \bar{V} = \frac{1}{\Nd}\sum_{i=1}^{\Nd} V_i \quad \text{and} \quad \sum_{i=1}^{\Nd} \delta V_i =0
	\;.
\end{align}
The dynamics of the volumes are given by Eq.~(6) in the main text.
To simplify notation, we introduce the auxiliary parameters $\hat{D} = D [ \frac{3}{4\pi} ]^{(1/3)}$ and $\hat{l}_\gamma = \ell_\gamma [\frac{4\pi}{3}]^{-(1/3)}$, %
\begin{align}
\frac{\mathrm{d} V_i}{\mathrm{d} t} = \frac{4\pi\hat{D}}{c_\mathrm{in}} \Big( V_i^\frac{1}{3}\Big(c_\infty-c_\mathrm{out} \Big) -\hat{l}_\gamma c_0 \Big)
\;,
\end{align}
where $c_\infty-c_\mathrm{out} = [(\bar{c}-c_\mathrm{out})V_\mathrm{sys} - (c_\mathrm{in}-c_\mathrm{out})\bar{V}]/[V_\mathrm{sys}-\bar{V}]$.
We start by analyzing the mean droplet volume~$\bar V$, whose dynamics are given by
\begin{align}
\frac{\mathrm{d} \bar V}{\mathrm{d} t}  =  \frac{4\pi\hat{D}}{c_\mathrm{in}} \frac{1}{\Nd}\left[
		\sum_{i=1}^{\Nd}(\bar{V}+\delta V_i)^{\frac{1}{3}} \bigl( c_\infty-c_\mathrm{out} \bigr) - \Nd\hat{l}_\gamma c_\mathrm{out}
	\right]
	\;.
\end{align}
Assuming all droplets are small ($\delta V_i \ll \bar V$), we expand to first order,
\begin{align}
	\sum_{i=1}^{\Nd}(\bar{V}+\delta V_i)^\frac{1}{3} \approx \Nd \bar{V}^\frac{1}{3}+\frac{1}{3}\bar{V}^{-\frac{2}{3}}\sum_{i=1}^{\Nd}\delta V_i  \approx \Nd \bar{V}^\frac{1}{3}
	\;,
\end{align}
so
 \begin{align}
	\frac{\mathrm{d} \bar V}{\mathrm{d} t} \approx  \frac{4\pi\hat{D}}{c_\mathrm{in}} \Big[ \bar{V}^\frac{1}{3} \Big( c_\infty-c_\mathrm{out} \Big) - \hat{l}_\gamma c_\mathrm{out} \Big]
	\label{eqn:vbar_dynamics}
	\;.
 \end{align} 
 For the $i$'th droplet, we thus have
 \begin{align}
	\frac{\mathrm{d} V_i}{\mathrm{d} t} &= \frac{\mathrm{d} \bar V}{\mathrm{d} t}  + \frac{\mathrm{d}\, \delta V_i}{\mathrm{d} t} = \frac{4\pi\hat{D}}{c_\mathrm{in}} \Big( (\bar{V}+\delta V_i)^\frac{1}{3}\bigl(c_\infty-c_\mathrm{out} \bigr) -\hat{l}_\gamma c_\mathrm{out} \Big)
\\& \approx \frac{4\pi\hat{D}}{c_\mathrm{in}} \Big( (\bar{V}^\frac{1}{3}+\frac{1}{3}\bar{V}^{-\frac{2}{3}} \delta V_i) \bigl(c_\infty-c_\mathrm{out} \bigr) -\hat{l}_\gamma c_\mathrm{out} \Big) 
\\
&\Rightarrow \frac{\mathrm{d}\, \delta V_i}{\mathrm{d} t} \approx \frac{4\pi\hat{D}}{c_\mathrm{in}}   \frac{\bar{V}^{-\frac{2}{3}}}{3}\delta V_i \Big(  c_\infty-c_\mathrm{out} \Big)
	\;.
 \end{align} 
To describe coarsening, we consider the relative droplet size, $\rho_i = \delta V_i/\bar{V}$, whose rate of change reads 
\begin{align}
	\frac{\mathrm d \rho_i}{\mathrm dt} = \left[
		\frac{\mathrm d \delta V_i}{\mathrm dt} \bar{V} 
		- \frac{\mathrm d \bar{V}}{\mathrm dt}\delta V_i
	\right]\bar{V}^{-2}
		\;.
\end{align}
Inserting the definitions of the time derivatives of $\bar{V}$ and $\delta V_i$ and expanding to first order gives
 \begin{align}
	\frac{\mathrm d \rho_i}{\mathrm dt}  = -\rho_i \frac{8\pi\hat{D}}{3c_\mathrm{in}}\cdot\frac{\bar{V}^\frac{1}{3} \Big(c_\infty-c_\mathrm{out} \Big) - \frac{3}{2}\hat{l}_\gamma c_\mathrm{out}}{\bar{V}} 
	\label{eqn:rho_dynamics}
	\;.
\end{align}
We next use this expression to analyze the dynamics of $\rho_i$ in the four different stages described in the main text.

We first consider the reaction stages.
For the stage $0 \to 1$, we have
\begin{align}
	V_{i}^{(1)} = V_{i}^{(0)}\Gamma^+  \quad \text{(Nucleoplasm} \xrightarrow{} \text{Droplet Material)} \label{eq:GammaPlus}
	\;.
\end{align}
Since each droplet increase by the same multiplicative factor $\Gamma^+$, $\rho_i$ remains constant during this stage. 
The same applies to the degradation stage $2 \to 3$, where $V_i^{(3)} = V_i^{(2)}\Gamma^-$. Applying the reaction protocol of the main text, $V_i^{(3)}-V_i^{(2)} = V_i^{(2)}\,(-2A)/(c_0+A)$, so that $\Gamma^- = (c_0-A)/(c_0+A)$. Because both reaction stages rescale every droplet by the same, identity-independent factor $\Gamma^\pm$, the relative volumes $\rho_i$ are unchanged during stages $0 \to 1$ and $2 \to 3$; coarsening can therefore only be modified by the diffusive transport in stages $1 \to 2$ and $3 \to 4$.

To study coarsening, we thus need to describe the evolution of $\rho_i$ from $1 \to 2$. 
If we let time run long enough for condensate material to equilibrate between the dilute phase and the dense phase, the average volume $\bar{V}$ will go to a steady state (SS) level, which is given by 
\begin{align}
	\bar{V}_\mathrm{SS}^+ =  \frac{1}{\Nd}\frac{c_0 + A -c_\mathrm{out}}{c_\mathrm{in}-c_\mathrm{out}} V_\mathrm{sys}
	\;.
\end{align}
To describe the approximate behavior of Eq.~\eqref{eqn:vbar_dynamics}, we next expand to first order around the volume $\bar{V}_\mathrm{SS}^+$.
To simplify notation, we express the nonlinear part of the equation as $f(\bar{V}) = \bar{V}^\frac{1}{3} [c_\infty(\bar{V})-c_\mathrm{out}]$, leading to
\begin{align}
	f(\bar{V}) &= f(\bar{V}_\mathrm{SS}^+) + f'(\bar{V}_\mathrm{SS}^+)(\bar{V}-\bar{V}_\mathrm{SS}^+) + \mathcal{O}([\bar{V}-\bar{V}_\mathrm{SS}^+]^2)
\\
	&\approx -\hat{c}(\bar{V}_\mathrm{SS}^+)^{-\frac{2}{3}}(\bar{V}-\bar{V}_\mathrm{SS}^+)
\\\notag
	& \text{ with } \quad \hat{c} = \frac{ (c_\mathrm{in}-c_\mathrm{out})(c_0+A-c_\mathrm{out})}{c_\mathrm{in}-(c_0+A)}
	\;.
\end{align}
Here, the first term in the first line, $ f(\bar{V}_\mathrm{SS}^+)$, is zero at equilibrium by definition.
Using this in Eq.~\eqref{eqn:vbar_dynamics},
\begin{align}
	\frac{\mathrm d \bar V}{\mathrm dt}  &\approx  \frac{4\pi\hat{D}}{c_\mathrm{in}} \Big[\hat{c}(\bar{V}_\mathrm{SS}^+)^{-\frac{2}{3}}(\bar{V}_\mathrm{SS}^+-\bar{V}) - \hat{l}_\gamma c_\mathrm{out} \Big]  
	\;,
\end{align}
we aim to solve for $\bar{V}$.
Introducing the auxiliary variables
\begin{align}
 \mu^+ &= \hat{c}\frac{4\pi\hat{D}}{c_\mathrm{in}} (\bar{V}_\mathrm{SS}^+)^{-\frac{2}{3}}
&\text{and}&&
 \psi^+ &= \hat{l}_\gamma \frac{c_\mathrm{out}}{\hat{c}}(\bar{V}_\mathrm{SS}^+)^{-\frac{1}{3}}   
 \;,
\end{align}
we can simplify,
\begin{align}
	\frac{\mathrm d \bar V}{\mathrm dt}  & \approx \mu^+ \Big[-\bar{V} + \bar{V}_\mathrm{SS}^+(1-\psi^+)\Big]
\end{align}
to find
\begin{align}
	\bar{V}(t) &\approx \bar{V}_\mathrm{SS}^+(1-\psi^+) + (\bar{V}_0 - \bar{V}_\mathrm{SS}^+(1-\psi^+)) e^{-\mu^+ t} 
	\;.
\end{align}
Inserting this result for dynamics of $\rho_i$ given by Eq.~\eqref{eqn:rho_dynamics}, we obtain
\begin{align}
	\frac{\mathrm d \rho_i}{\mathrm dt} & \approx \rho_i \mu^+\frac{2}{3}  \Big[1 -  \frac{\bar{V}_\mathrm{SS}^+(1-\frac{3}{2}\psi^+) }{\bar{V}} \Big ]
	\\\Rightarrow
	\rho_i(t) &= \rho_i(0)e^{\frac{2}{3}\mu^+ [t - \bar{V}_\mathrm{SS}^+(1-\frac{3}{2}\psi^+)\int \frac{1}{\bar{V}}\mathrm{d}t}K
	\;.
\end{align}
Where K is the integration constant, to be determined from the initial condition. Since $\int \bar{V}^{-1}\mathrm{d}t = \frac{\ln(\bar{V})+\mu^+t}{\mu^+\bar{V}_\mathrm{SS}^+(1-\psi^+)}$, we end up with 
\begin{align}
	\rho_i (t) = e^{\frac{2}{3}\mu^+t \Big(1-\frac{1-\frac{3}{2}\psi^+}{1-\psi^+}\Big)}\Big( \frac{\bar{V}(t)}{\bar{V}^{(1)}}\Big) ^{-\frac{2}{3}\Big(\frac{1-\frac{3}{2}\psi^+}{1-\psi^+}\Big)}
	\;.
\end{align}
Since both $\ell_\gamma$ and $c_\mathrm{out}/\hat{c}$ are small numbers, $\psi^+$ is small ($\psi^+ \ll 1$), so $1-\frac{1-\frac{3}{2}\psi^+}{1-\psi^+} \approx \frac{\psi^+}{2}$, giving 
\begin{align}
	\rho_i^{(2)} &= \rho_i^{(0)} e^{\frac{1}{3}\mu^+\psi^+\frac{\pi}{\omega}} \Big( \frac{\bar{V}^{(2)}}{\bar{V}^{(0)}\Gamma^+}\Big) ^{-\frac{2}{3}}
	&\text{with}\quad \mu^+\psi^+ &= \frac{4\pi D \ell_\gamma \Nd}{V_\mathrm{sys}}\frac{c_\mathrm{out}}{c_\mathrm{in}}\frac{c_\mathrm{in}-c_\mathrm{out}}{c_0 + A - c_\mathrm{out}}
	\;.
\end{align}
This is equivalent with Eq.~(8) in the main text. Note that $\rho_i^{(2)}$ and $\bar{V}^{(2)}$ are equivalent to $\rho_i(t=\frac{\pi}{\omega})$ and $\bar{V}(t=\frac{\pi}{\omega})$, respectively, which is why $\omega$ enters in the exponential. 
To simplify notation, we introduce the amplitude independent rate parameter $\Psi$,
\begin{align}
	\Psi &= \lambda \frac{c_\mathrm{in}-c_\mathrm{out}}{c_0 - c_\mathrm{out}}
	& \text{with} &&
	\lambda &= \frac{4\pi D \ell_\gamma \Nd}{V_\mathrm{sys}}\frac{c_\mathrm{out}}{c_\mathrm{in}}
\;,
\end{align}
where $\Psi$ is the constant used in the main text.
This allow us to estimate $\rho_i$ after a full cycle, using
\begin{align}
	V_{i}^{(3)} = V_{i}^{(2)}\Gamma^- \quad \text{(Droplet Material} \xrightarrow{} \text{Nucleoplasm)}
	\;,
\end{align}
with which we obtain
\begin{align}
	\rho_i^{(4)} &= \rho_i^{(2)}e^{\frac{1}{3}\mu^-\psi^-\frac{\pi}{\omega}} \Big( \frac{\bar{V}^{(4)}}{\bar{V}^{(3)}}\Big) ^{-\frac{2}{3}}
\\
	&= \rho_i^{(0)}e^{\frac{1}{3}\lambda(\frac{c_\mathrm{in}-c_\mathrm{out}}{c_0+A-c_\mathrm{out}} + \frac{c_\mathrm{in}-c_\mathrm{out}}{c_0-A-c_\mathrm{out}})\frac{\pi}{\omega}}\Big( \frac{\bar{V}^{(4)}}{\bar{V}^{(2)}\Gamma^-}\frac{\bar{V}^{(2)}}{\bar{V}^{(0)}\Gamma^+}\Big) ^{-\frac{2}{3}}
	\;.
\end{align}
Assuming the amount of material inside the droplets over a full cycle is in steady state ($\bar{V}^{(0)} = \bar{V}^{(4)}$), the last parenthesis reduce to $(\Gamma^-\Gamma^+)^{2/3}$.
Further assuming amplitudes are small compared to the mean value $c_0$ (by definition they have to be smaller), so $(c_0-c_\mathrm{out})^2 \gg A^2$, we have after a full cycle
\begin{align}
	\frac{\rho_i^{(4)}}{\rho_i^{(0)}} \approx e^{\frac{1}{3} \lambda \frac{c_\mathrm{in}-c_\mathrm{out}}{c_0 - c_\mathrm{out}}\frac{2\pi}{\omega}} \Big(\Gamma^-\Gamma^+\Big)^{2/3}
	= e^{\frac{1}{3}\frac{4\pi D\ell_\gamma}{\bar{V}^0_\mathrm{SS}}\frac{c_\mathrm{out}}{c_\mathrm{in}}\frac{2\pi}{\omega}} \Big(\Gamma^-\Gamma^+\Big)^{2/3} 
	\;.
\end{align}
On the right side, we have applied the steady state average droplet volume $\bar{V}^0_\mathrm{SS} = \frac{1}{\Nd}\frac{c_0 -c_\mathrm{out}}{c_\mathrm{in}-c_\mathrm{out}} V_\mathrm{sys}$ since this might be a direct experimental observable.
To estimate the relation between $A$ and $\omega$ ensuring stability, we note that each $\rho_i$ is independent of all others, indicating that stability is global for all $\rho_i$, so we can use $\rho_i=\rho$.
For stability, we require $\rho (t=\frac{2\pi}{\omega})/\rho (t=0) < 1$, so
\begin{align}
	1 &= e^{\frac{1}{3} \lambda \frac{c_\mathrm{in}-c_\mathrm{out}}{c_0 - c_\mathrm{out}}\frac{2\pi}{\omega}} \Big(\Gamma^-\Gamma^+\Big)^{2/3}
\\
	\Rightarrow -\ln\Big(\Gamma^-\Gamma^+\Big) &= \lambda \frac{c_\mathrm{in}-c_\mathrm{out}}{c_0 - c_\mathrm{out}}\frac{\pi}{\omega}
	\;.
\end{align}
Applying $\Gamma^+ = 1$ and $\Gamma^- = 1 - \frac{2A}{c_0+A}$, and approximating $A \ll c_0$, we obtain the stability criteria
\begin{align}
	A\omega >\frac{\pi}{2}\lambda\frac{c_\mathrm{in}-c_\mathrm{out}}{c_0 - c_\mathrm{out}}c_0 =  2\pi \frac{D\ell_\gamma}{V_{SS}^0}\frac{c_\mathrm{out}}{c_\mathrm{in}}c_0 = \frac{\pi}{2}\Psi c_0
	\;,
\end{align}
which is identical to \Eqref{eq:omegaArelation} in the main text.

\section{Stability criterion for sinusoidal wave forms}
\label{sec:appendix_sine}
Finally, we extend the analysis above to sinusoidal forcing by considering the limit of infinitesimal additions and removals.
This yields an expression for $\rho_i$ under continuous driving, recovering the same stability criterion as in the square-wave case in the small-amplitude limit.
To do this, we divided the time period $t\in[0, T]$ with $T=2\pi/\omega$ into $n$ small steps of duration $\Delta t$ and express the dynamics of $\rho$ iteratively,
\begin{align}
	\rho_{n+1} &= \rho_{n} e^{\frac{\lambda}{3}\frac{c_\mathrm{in}-c_\mathrm{out}}{\bar{c}(t_n) - c_\mathrm{out}}\Delta t}\Big( \frac{\bar{V}_{n+1}}{\bar{V}_{n}\Gamma^\pm}\Big) ^{-\frac{2}{3}}
&
	\text{ with } && \Gamma_n^\pm &= \begin{cases}\Gamma^+ & \bar{c}'(t_n) > 0\;, \\[6pt]
\Gamma^- & \bar{c}'(t_n) \leq 0\;.
\end{cases}
\end{align}
After $N=nT$ steps, we have
\begin{align}
\frac{\rho_N}{\rho_0}
= \prod_{n=0}^{N-1}
e^{\frac{\lambda}{3}\frac{c_\mathrm{in}-c_\mathrm{out}}{\bar{c}(t_n) - c_\mathrm{out}}\Delta t}\;
\Big(\frac{\bar{V}_n}{\bar{V}_{n-1}}\;\frac{1}{\Gamma_n^\pm}\Big)^{-\frac{2}{3}}.
\end{align}
Passing to the continuum limit $\Delta t\to 0$, we apply Volterra product integral, and find the contribution of each factor,
\begin{subequations}
	\begin{align}
		\prod_{n=0}^{N-1}e^{\frac{\lambda}{3}\frac{c_\mathrm{in}-c_\mathrm{out}}{\bar{c}(t_n) - c_\mathrm{out}}\Delta t}
		&\;=\;
		e^{ \frac{\lambda}{3}(c_\mathrm{in}-c_\mathrm{out}) \int_0^\frac{2\pi}{\omega} \frac{\mathrm{d}t}{\bar{c}(t)-c_\mathrm{out}} },\\[6pt]
		\prod_{n=0}^{N-1}\frac{\bar{V}_n}{\bar{V}_{n-1}}
		&\;=\;\frac{\bar{V}_N}{\bar{V}_0}\;=\;1,
		\\[6pt]
		\prod_{n=0}^{N-1}(\Gamma_n^\pm)^{2/3}
		&\;=\;
		e^{ \frac{2}{3} \int_{\pi/2}^{3\pi/2} \frac{\bar{c}'(t)}{\bar{c}(t)} \mathrm{d}t }
		\label{eqn:integral3}
		\;,
	\end{align}
\end{subequations}
where the last line follows from the fact that steps with $c'(t)>0$ contribute $1$ since $\Gamma^+ = 1$. Solving the integral in the first line for $\bar{c}(t) = c_0 + A\sin(\omega t)$ gives
\begin{align}
	\int_0^\frac{2\pi}{\omega} \frac{\mathrm{d}t}{c_0-c_\mathrm{out} + A\sin(\omega t)}
	&= \frac{2\pi}{\omega\,\sqrt{(c_0-c_\mathrm{out})^2-A^2}},
	\qquad \text{for}\quad|c_0-c_\mathrm{out}|>|A|
	\;.
\end{align}
Second, the integral in Eq.~\eqref{eqn:integral3} for $c'(t)<0$ gives
\begin{align}
\int_{\frac{\pi}{2\omega}}^{\frac{3\pi}{2\omega}} \frac{A\omega\cos(\omega t)}{c_0 + A\sin(\omega t)}\,\mathrm{d}t
= \ln\!\left(\frac{c_0 - A}{c_0 + A}\right)
	\;.
\end{align}
Combining all factors, we obtain
\begin{align}
	\frac{\rho^{(4)}}{\rho^{(0)}} = e^{\frac13\Psi\frac{2\pi}{\omega}} \Big( \frac{c_0-A}{c_0+A} \Big)^{2/3} 
	\;.
\end{align}
Expanding both terms (inside the bracket) to first order (assuming small $A$), we obtain
\begin{align}
	\frac{\rho^{(4)}}{\rho^{(0)}}\approx e^{\frac13\Psi\frac{2\pi}{\omega}} \Big( 1-\frac{2A}{c_0}\Big)^{2/3} = e^{\frac13\Psi\frac{2\pi}{\omega}} \Big(\Gamma^+\Gamma^-  \Big)^{2/3}
	\;.
\end{align}
Since this is the same expression as for the square wave (\Eqref{eq:Rhostabilization} in the main text), we find the same stability condition (given by \Eqref{eq:omegaArelation} in the main text).

\end{document}